\documentclass[prx,groupedaddress,showpacs,preprintnumbers,amsmath,amssymb,twocolumn,floats]{revtex4}
\usepackage{txfonts}
\usepackage{amssymb}
\usepackage{graphicx}
\usepackage{color}

\begin{document}

\title{Electronic Structure Reconstruction across the Antiferromagnetic Transition in TaFe$_{1.23}$Te$_3$ Spin Ladder}

\author{M. Xu$^{1}$}
\author{Li-Min Wang$^{2}$}
\author{R. Peng$^{1}$}
\author{Q. Q. Ge$^{1}$}
\author{F. Chen$^{1}$}
\author{Z. R. Ye$^{1}$}
\author{Y. Zhang$^{1}$}
\author{S. D. Chen$^{1}$}
\author{M. Xia$^{1}$}
\author{R. H. Liu$^{3}$}
\author{M. Arita$^{4}$}
\author{K. Shimada$^{4}$}
\author{H. Namatame$^{4}$}
\author{M. Taniguchi$^{4}$}
\author{M. Matsunami$^{5}$}
\author{S. Kimura$^{5}$}
\author{M. Shi$^{6}$}
\author{Wei Ku$^{2}$}
\author{X. H. Chen$^{3}$}
\author{Wei-Guo Yin$^{2}$}\email{wyin@bnl.gov}
\author{B. P. Xie$^{1}$}\email{bpxie@fudan.edu.cn}
\author{D. L. Feng$^{1}$}\email{dlfeng@fudan.edu.cn}

\affiliation{$^{1}$State Key Laboratory of Surface Physics, Department of Physics, and Advanced Materials Laboratory, Fudan University, Shanghai 200433, People's Republic of China}

\affiliation{$^{2}$Condensed Matter Physics and Materials Science Department, Brookhaven National Laboratory, Upton, New York 11973, USA}

\affiliation{$^{3}$Hefei National Laboratory for Physical Sciences at Microscale and Department of Physics, University of Science and Technology of China, Hefei, Anhui 230026, China}

\affiliation{$^{4}$Hiroshima Synchrotron Radiation Center and Graduate School of Science, Hiroshima University, Hiroshima 739-8526, Japan.}

\affiliation{$^{5}$UVSOR Facility, Institute for Molecular Science and The Graduate University for Advanced Studies, Okazaki 444-8585, Japan}

\affiliation{$^{6}$Swiss Light Source, Paul-Scherrer Institut, 5232 Villigen, Switzerland}

\begin{abstract}
With angle-resolved photoemission spectroscopy, we studied the electronic structure of TaFe$_{1.23}$Te$_3$, which is a two-leg spin ladder compound with a novel antiferromagnetic ground state.
Quasi-two-dimensional Fermi surface is observed, indicating sizable inter-ladder hopping, which would facilitate the in-plane ferromagnetic ordering through double exchange interactions.
Moreover,  an energy gap is not observed at the Fermi surface in the antiferromagnetic state. Instead, the shifts of various bands  have been observed.
Combining these observations with density-functional-theory calculations, we propose that the large scale reconstruction of the electronic structure, caused by the interactions between the coexisting itinerant electrons and local moments, is most likely the driving force behind the magnetic transition.
TaFe$_{1.23}$Te$_3$ thus provides a simpler system that contains similar ingredients as the parent compounds of iron-based superconductors, which yet could be readily modeled and understood.
\end{abstract}


\maketitle

\section{Introduction}

High-temperature superconductors discovered so far are always quasi-two-dimensional (2D) systems in the vicinity of a  magnetic order \cite{cu1, Fe1, Fe2}. However, due to many body interactions, the analytical solutions of related theoretical models in two dimensions are often hard to obtain. For cuprate, one alternative way to attack the problem is the following successive steps:  firstly solving a one-dimensional chain, then a two-leg ladder consisted of two interacting chains, and finally extending the solution to an N-leg ladder system \cite{curpate1}. In the N$\rightarrow\infty$ limit, similar physics and phenomena in 2D systems can be asymptotically approached. Moreover,  chains or ladders often contain the same physical ingredients and exhibit similar properties  as  2D systems. For example, superconductivity has been observed in doped the cuprate spin ladders, \textit{eg.} Sr$_2$Ca$_{12}$Cu$_{24}$O$_{41}$ \cite{curpate2, curpate3}. The mechanisms of novel pheonomena in 2D systems thus might be demonstrated in chains or ladders with greater clarity due to the lower dimensionality.

\begin{figure}[t]
\includegraphics[width=8.6cm]{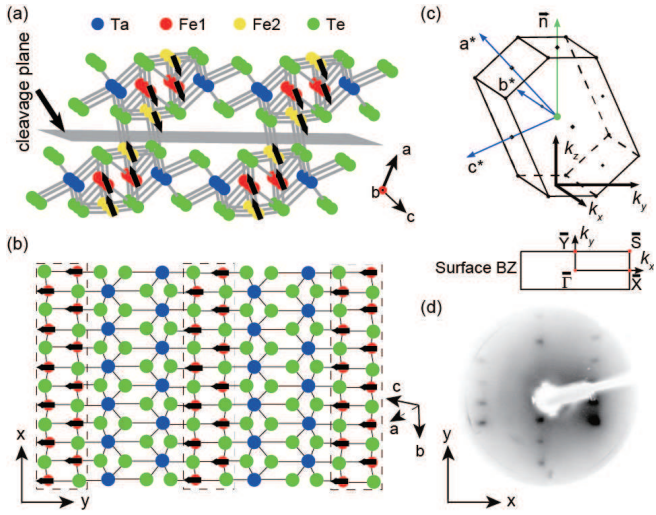}
\caption{(Color online). Crystal structure and Brillouin zone (BZ) of TaFe$_{1.23}$Te$_3$. (a) Schematic illustration of crystal structure and spin structure following neutron scattering result in Ref. \onlinecite{mao neutron}. The Fe2 atoms partially and randomly occupy the interstitial sites. (b) A projection of the ladder structure onto the natural cleavage (-101) plane. The Fe-Fe zig-zag two-leg ladders are encircled by dashed lines. Note that the interstitial Fe2 is not shown here. (c) Three-dimensional BZ of TaFe$_{1.23}$Te$_3$ with the monoclinic structure. The direction of vector $\vec{n}$ is normal to the cleaved surface (-101) in the reciprocal space. The bottom part gives the surface BZ. Hereafter, we define the $\bar{\Gamma}$$\bar{X}$ and $\bar{\Gamma}$$\bar{Y}$ axis to be parallel to and perpendicular to the  ladder, respectively. $\bar{\Gamma}$$\bar{X}$= 0.863~${\AA}^{-1}$ and $\bar{\Gamma}$$\bar{Y}$= 0.305~${\AA}^{-1}$. (d) The low-energy electron diffraction (LEED) pattern of TaFe$_{1.23}$Te$_3$ taken at 250~K in the paramagnetic state with 100~eV incident electrons. Note that $\bar{\Gamma}$$\bar{X}$/$\bar{\Gamma}$$\bar{Y} \approx$2.8 is consistence with the LEED pattern.} \label{intro}
\end{figure}

Analogously, magnetism and superconductivity in iron-based compounds  could be studied in a similar way \cite{Dagotto}. Several spin chain/ladder systems have been discovered experimentally, including TaFe$_{1.23}$Te$_3$ \cite{liusample}, BaFe$_2$Se$_3$ \cite{leiBa, BaFeSe}, Ce$_2$O$_2$FeSe$_2$ \cite{CeFeSe}, and the single layer K$_x$Fe$_{2-y}$Se$_2$ (110) film \cite{liFeSe}.
Intriguingly, the signature of superconductivity has been observed in an alkali-doped FeSe-ladder system \cite{liFeSe}, and the antiferromagnetic (AFM) order similar to the spin density wave (SDW)  in the parent compounds of iron-based superconductors has been observed as well \cite{CeFeSe}. Meanwhile, all these materials share a similar structural character: a layered, quasi-one-dimensional structure with edge sharing FeX$_4$ (X=Te, Se) plaquettes. For example, TaFe$_{1.23}$Te$_3$ can be viewed as Fe-Fe zig-zag spin ladder compound with adjacent ladders connected by a Ta/Te network as illustrated in Figs.~\ref{intro}(a) and \ref{intro}(b). It exhibits an AFM order with an unusual ferromagnetic (FM) coupling in the cleavage plane and an antiferromagnetic coupling out of plane as shown in Fig.~\ref{intro}(a), and the local moment is about 2~$\mu_B$/Fe, which resembles that of FeTe \cite{mao neutron, DaiFeTe}.
Due to their similarities, these ladder compounds can be considered as low dimensional siblings of the iron-based superconductors. Besides the interests in themselves as iron-based spin ladders, experimental study  in combination with readily treatable theoretical modeling of these systems would deepen our understanding of magnetism in iron-based superconductors.

In this paper, we study the electronic structure of TaFe$_{1.23}$Te$_3$  by angle-resolved photoemission spectroscopy (ARPES) and band calculations. We have observed quasi 2D Fermi surface  in TaFe$_{1.23}$Te$_3$.  An energy  gap is not observed on the Fermi surface, indicating  the absence of  Fermi surface instability. Instead, we find that  features far below the Fermi energy ($E_F$) shift abruptly across the AFM transition, resembling the electronic structure reconstruction across SDW transition in 2D iron pnictides \cite{zhangNaFeAs, Yang, Ge, YiBaCo}. Our results suggest that the critical ingredients of the magnetism in the parent compounds of iron-based superconductors, such as local moments and the Hund's rule coupling, together with the double exchange interactions across the in-plane ladders, conspire the novel ground state in   TaFe$_{1.23}$Te$_3$.
It is thus another new compound whose order is characterized by electronic structure reconstruction instead of Fermi surface nesting, and provides a simpler playground to investigate the magnetism in iron-based superconductors and related materials.

\section{Experimental Procedures}

\begin{figure}
\includegraphics[width=8.6cm]{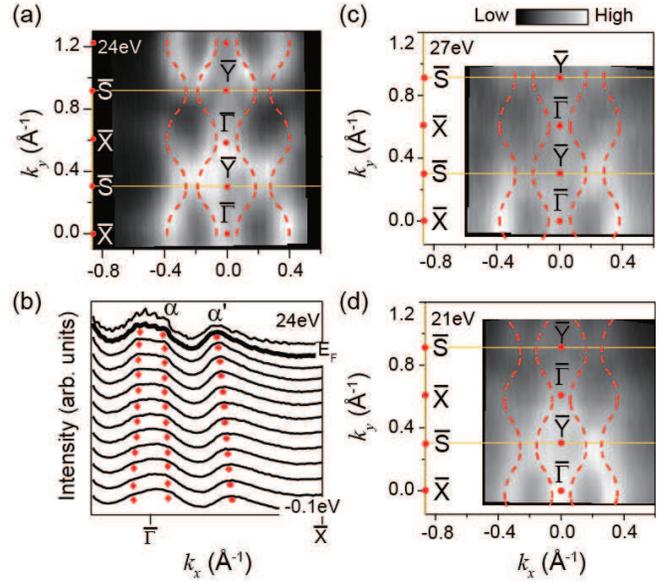}
\caption{(Color online). The Fermi surface of TaFe$_{1.23}$Te$_3$. (a) Photoemission intensity maps integrated within 10~meV around Fermi energy ($E_F$) with 24~eV photons at 30~K. (b) Momentum distribution curves (MDC's) near $E_F$ along $\bar{\Gamma}\bar{X}$. Marks are guide to eyes of the  dispersions. (c) and (d) Photoemission intensity maps taken with 27, and 21~eV photons, respectively. The intensity was integrated within 10~meV around $E_F$. The red dashed lines are the guides to eyes for the Fermi surface, which show weak $k_z$ dependence. Data were taken with horizontally polarized photons at 13~K at UVSOR.} \label{FS}
\end{figure}

TaFe$_{1.23}$Te$_3$ single crystals were synthesized by chemical vapor transport method as described elsewhere \cite{liusample}, which show flat shiny and needle-shaped surfaces. With decreasing temperature, the resistivity shows metallic behavior and has an abnormal transition at the Neel temperature ($T_N$) of about 160~K \cite{liusample}, resembling the resistivity anomaly of BaFe$_2$As$_2$ at its SDW transition \cite{BaFeAs}. The chemical compositions were determined by energy dispersive X-ray (EDX) spectroscopy. As illustrated in  Fig.~\ref{intro}(a), TaFe$_{1.23}$Te$_3$ is in a monoclinic P2$_1$/m structure with lattice constants a =7.4262~$\AA$, b =3.6374~$\AA$, c = 9.9925~$\AA$, and $\beta$ =~$109.166\textordmasculine $. On the (-101) natural cleavage plane  schematically plotted in Fig.~\ref{intro}(b), there are two-leg ladders with two FeTe chains parallel to the short axis b, which are separated by a Ta/Te network in-between. Therefore, TaFe$_{1.23}$Te$_3$ possesses a quasi-one-dimensional crystal structure.
 Fig.~\ref{intro}(c) illustrates  the three-dimensional Brillouin zone (BZ), and the surface BZ. We define the $k_x$ and $k_y$ axes  to be  parallel  and perpendicular to the ladder/chain, respectively. The high-quality sample surface was confirmed by the clear pattern of low-energy electron diffraction (LEED) as shown in Fig.~\ref{intro}(d), which manifests its twofold symmetry.

The ARPES experiments were performed with circularly polarized synchrotron light from beamline 9A of Hiroshima Synchrotron Radiation Center (HSRC), the SIS beamline of the Swiss Light Source (SLS) with Scienta R4000 electron analyzers, and BL7 of Ultraviolet Synchrotron Orbital Radiation (UVSOR) facility with an MBS A-1 electron analyzer. The overall energy resolution is $\sim15$~meV, and angular resolution is $\sim0.3$~degree. The samples were cleaved \textit{in situ}, and measured under ultra-high-vacuum of~$5\times10^{-11}~$\textit{Torr}.

\section{Results}
\subsection{Fermi surface and band structure}
\begin{figure}[t]
\includegraphics[width=8cm]{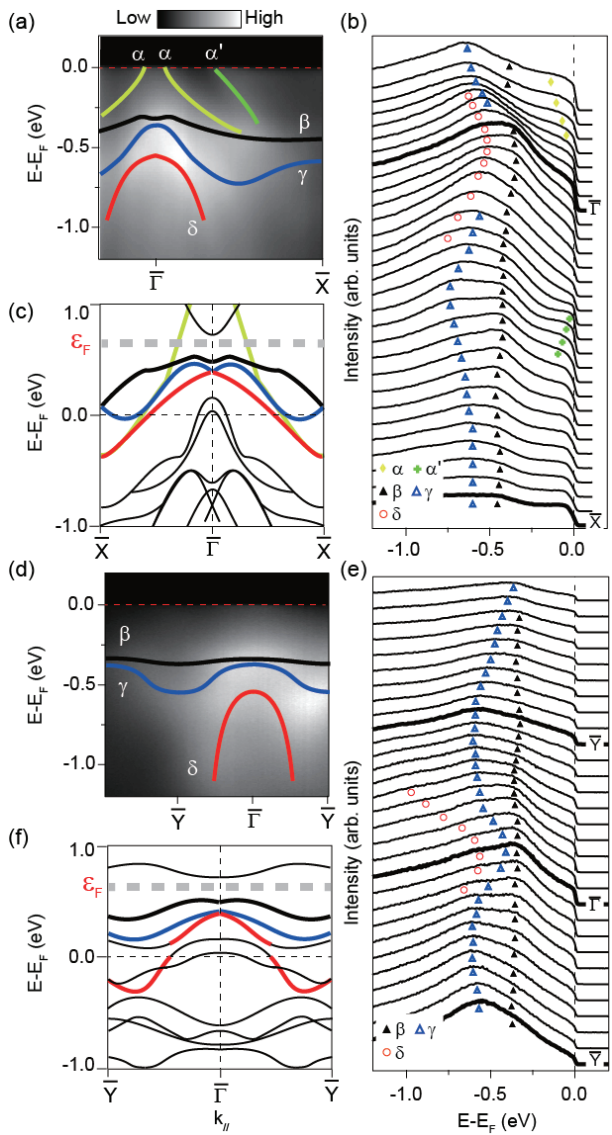}
\caption{(Color online) Band structure of TaFe$_{1.23}$Te$_3$. (a) Photoemission intensity along $\bar{\Gamma}\bar{X}$ which is parallel to the ladder. (b) Plot of energy distribution curves (EDC's) along $\bar{\Gamma}\bar{X}$ in panel (a). (c) The calculated band structure along $\bar{\Gamma}\bar{X}$ without interstitial iron atoms Fe$_2$, the thickened part of the bands have been  observed in our data. (d)-(f) are the same as panels (a)-(c), but for data along $\bar{\Gamma}\bar{Y}$ which is perpendicular to the Fe-Fe zig-zag ladder.  Marks are guide to eyes of the band dispersion which are determined by the local maxima of the EDC's. Data in panels (a) and (b) were taken with horizontally polarized 24~eV photons at 13~K at UVSOR. Data in panels (d) and (e) were taken with circularly polarized 24~eV photons at 10~K at SLS. The thick dashed lines in panels (c) and (f) indicate where the experimental Fermi energy is situated. } \label{cut}
\end{figure}

 According to the photoemission intensity map of TaFe$_{1.23}$Te$_3$ shown in Fig.~\ref{FS}(a), one could observe four Fermi surface sheets. Meanwhile, the periodic undulation-like Fermi surface along $\bar{\Gamma}\bar{Y}$ suggests sizable in-plane inter-ladder interactions. Figure~\ref{FS}(b) plots the momentum distribution curves (MDC's) along $\bar{\Gamma}\bar{X}$, and two bands $\alpha$ and ${\alpha}^{\prime}$ could be observed near $E_F$.  The Fermi surface of ${\alpha}^{\prime}$ seems to have the same shape as that of  $\alpha$, but judging from their dispersions, ${\alpha}^{\prime}$  is not a folded band of $\alpha$.
More detailed study of the \textit{k$_z$} dependence of the Fermi surface has been conducted with several photon energies, as shown in Figs~\ref{FS}(c) and \ref{FS}(d). Again, one can use almost the same Fermi surface sheets as those in Fig.~\ref{FS}(a) to match all the data at different \textit{k$_z$'s}. The weak $k_z$ dependence of Fermi surface  demonstrates the weak inter-plane coupling or quasi 2D nature of TaFe$_{1.23}$Te$_3$.

  Figure~\ref{cut}(a) displays the photoemission intensity along $\bar{\Gamma}\bar{X}$ (along the ladder), and the corresponding energy distribution curves (EDC's) are plotted in Fig.~\ref{cut}(b). Besides the $\alpha$ and ${\alpha}^{\prime}$ bands near $E_F$, three other bands $\beta$, $\gamma$, and $\delta$ can be identified. The $\beta$ and $\gamma$ bands disperse from 0.3 to 0.7~eV below $E_F$, while $\delta$  shows a parabola-like dispersion with a larger bandwidth.  Figures~\ref{cut}(d) and~\ref{cut}(e) display the photoemission data and the corresponding EDC's along $\bar{\Gamma}\bar{Y}$ (perpendicular to the ladder). One could only observe the $\beta$, $\gamma$, and $\delta$ bands. We note that for FeTe, coherent quasi-particles show up in its bicollinear antiferromagnetic  state, when the magnetic fluctuations are gapped out \cite{zhangFeTe, liuFeTe}. But sharp quasi-particle peak is not observed here in TaFe$_{1.23}$Te$_3$, which may be attributed to stronger magnetic fluctuations in ladders.

We calculated the band structure of TaFeTe$_3$ in the non-magnetic state without the interstitial Fe2 atoms. The calculation is conducted using the WIEN2K implementation of the full potential linearized augmented plane wave method in the local density approximation \cite{calc1}. The \textit{k}-point mesh was taken to be 4$\times$5$\times$11. The lattice constants were taken from ref.~\onlinecite{calc2}. To compare with the APRES spectra directly, the band structure in 2-Fe zone is unfolded to the 1-Fe zone by applying the recent developed unfolding method \cite{calc3}. Figures~\ref{cut}(c) and \ref{cut}(f) exhibit the corresponding calculated band dispersions along $\bar{\Gamma}$$\bar{X}$ and $\bar{\Gamma}$$\bar{Y}$, respectively. The thickened part of the bands in the calculation are likely the  bands observed in the experiment. The calculations qualitatively agree with the experiments, when considering some moderate band renormalization and the Fermi level shift due to the interstitial iron atoms in the real material (shown by the thick dashed lines). Note that the ${\alpha}^{\prime}$ band is absent in our calculations, and its origin is still mysterious. One possibility is that ${\alpha}^{\prime}$ may be due to the interstitial iron atoms which are not included in the calculations.

 \begin{figure*}
\includegraphics[width=17.2cm]{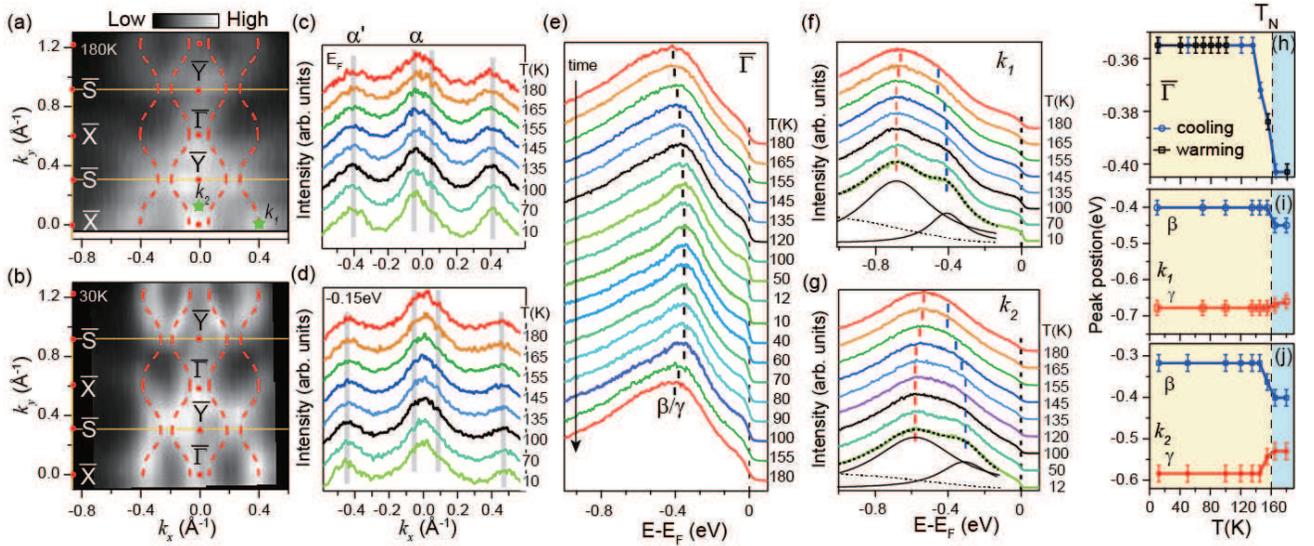}
\caption{(Color online). The temperature dependence of Fermi surface and electronic structure of TaFe$_{1.23}$Te$_3$. (a) and (b) Photoemission intensity map integrated within 10~meV around $E_F$, in paramagnetic (180~K) and antiferromagnetic (30~K) states, respectively. (c) Temperature dependence of MDC's at $E_F$ along $\bar{\Gamma}\bar{X}$. (d) is the same as panel (c) but for data chosen at 0.15~eV below $E_F$. (e) Temperature cycling data of EDC's at $\bar{\Gamma}$. The bars illustrate the peak positions of the EDC's. (f) Temperature dependence of EDC's at a selected momentum $k_{1}$ marked in panel (a). (g) is the same as panel (f) but for data at $k_{2}$ marked in panel (a). In panels (f) and (g), two examples of the fitted curves, Shirley background and Lorentz functions are shown with thin dashed lines and solid lines, respectively. (h)-(j) The summary of the corresponding peak position of the EDC's in panels (e)-(g), respectively. Data in panels (a) and (b) were taken with horizontally polarized 24~eV photons at UVSOR. The temperature dependence data in panels (c)-(j) were taken with circularly polarized 24~eV photons at HSRC.} \label{TD}
\end{figure*}

\begin{figure}
\includegraphics[width=\columnwidth]{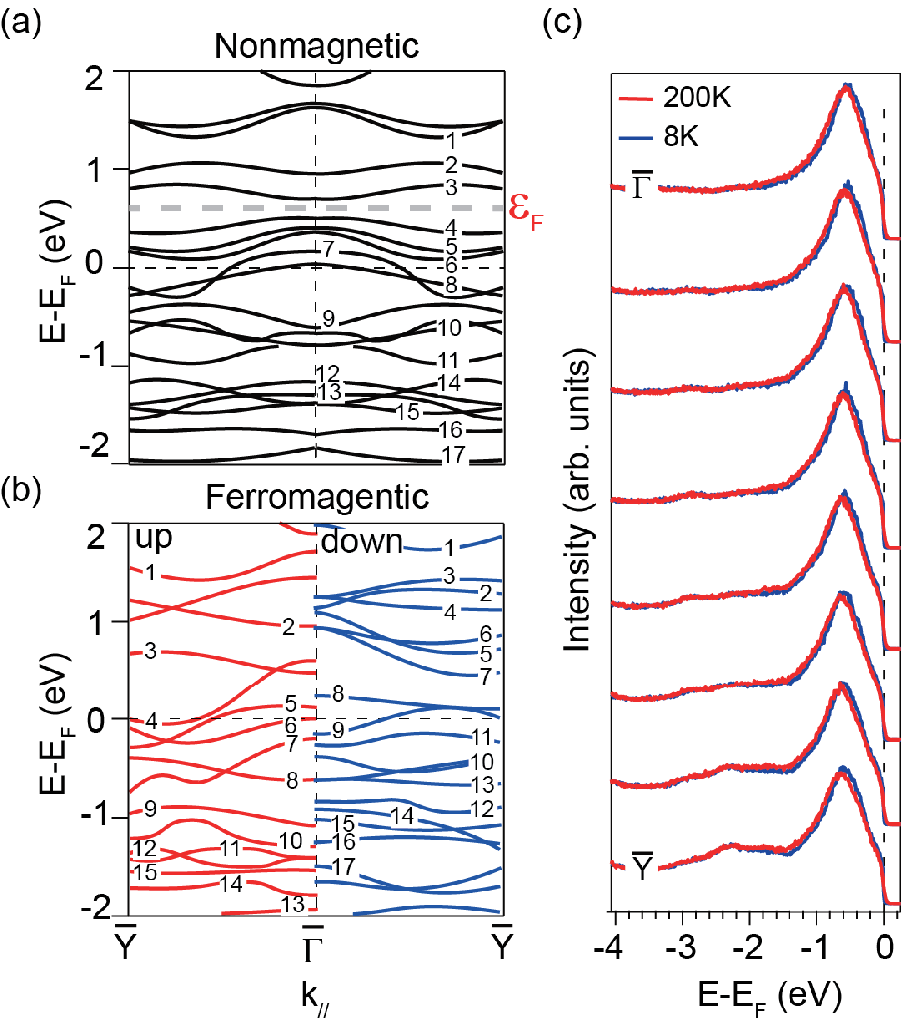}
\caption{(Color online).(a) and (b) The calculated valance band along $\bar{\Gamma} \bar{Y}$ of TaFeTe$_3$ for the nonmagnetic state, and the ferromagnetic spin-up (spin-majority) and spin-down (spin-minority) states, respectively. The band indices 1-17 are the guide to eyes. The thick dashed line in panel (a) indicates where the experimental Fermi energy is situated.  (c) The comparison of valance band along $\bar{\Gamma} \bar{Y}$ of TaFe$_{1.23}$Te$_3$ at 8~K and 200~K. Here data were taken with an in-house helium discharge lamp (21.2~eV).} \label{VB}
\end{figure}

\subsection{Temperature dependence of the electronic structure}

Temperature dependence of the spectra were taken to investigate the  mechanism of the AFM transition in TaFe$_{1.23}$Te$_3$. Figures~\ref{TD}(a) and~\ref{TD}(b) plot the Fermi surface intensity maps in the paramagnetic and AFM state, respectively. The Fermi surface exhibits negligible change except some thermal broadening, and there is no SDW gap at the Fermi surface. Figures~\ref{TD}(c) and~\ref{TD}(d) display temperature dependence of the MDC's along $\bar{\Gamma}\bar{X}$, which show that the movements of $\alpha$ and ${\alpha}^{\prime}$ are negligible as well. These results suggest that the AFM transition is not driven by the Fermi surface instability, although  the AFM coupling is out-of-plane,   the nesting condition would be fullfilled due to the weak $k_z$ dependence.
On the other hand, Fig.~\ref{TD}(e) shows the temperature cycling data of EDC's at $\bar{\Gamma}$, where the feature near -0.4~eV is contributed by the $\beta$ and $\gamma$ bands, and its temperature dependence of corresponding peak is summarized in Fig.~\ref{TD}(h). It clearly shows that $\beta$/$\gamma$  shifts towards lower binding energy as temperature decreases; when warming up, the peak shifts back again without any observable hysteresis, indicating the second order nature of the transition.
Figures~\ref{TD}(f) and \ref{TD}(g) display the temperature dependence of EDC's across the AFM transition at two representative momenta $k_{1}$ and $k_{2}$  respectively, as marked in  Fig.~\ref{TD}(a). As shown in Figs.~\ref{TD}(h)-\ref{TD}(j), the shifts of their peak positions all begin around the bulk $T_N$, and saturate quickly below 145~K  \cite{liusample}. Because of the absence of structural transition at $T_N$ in TaFe$_{1.23}$Te$_3$, such an intrinsic electronic structure reconstruction should be related to the AFM transition. It also indicates that our photoemssion data  should  reflect the bulk properites.
The amplitude of the shift is about 50~meV upwards for $\beta$,  18~meV downwards for $\gamma$ at $k_1$ in Fig.~\ref{TD}(i), and 80~meV upwards for $\beta$, 50~meV downwards for $\gamma$ at $k_2$ in Fig.~\ref{TD}(j). These are comparable with the $k_B T_N$ energy scale and those band shifts observed  in the parent compounds of the iron-based superconductors \cite{Yang, YiBaCo}. We note that the upward shifts  are larger than the downward shifts, and it seems the electronic energy is not saved here, different from the observation in 2D iron pnictides \cite{zhangNaFeAs,Yang} and FeTe \cite{zhangFeTe}.

Considering the measured quasi-2D nature of the  electronic structure and the in-plane FM spin alignment, the electronic structure in the AFM state may be represented by calculating the FM state in a single plane. Figs.~\ref{VB}(a) and  \ref{VB}(b) show the valence bands calculated  for the non-magnetic and FM state of  TaFeTe$_3$, respectively, which demonstrate a significant electronic structure reconstruction.  The calculated band shifts are in the order of $0.5 \sim 1$~eV, much larger than the experimental ones. However, it does indicate that the energy saving through the large scale electronic structure reconstruction is most likely the driving force behind the magnetic transition.
Experimentally, the Fermi energy is in between  band 3 and band 4 of the normal state band structure, since Fe2 in the real matrials would bring in more electrons, which were not considered in the calculation.  In the AFM state, the spin down bands and spin up bands generally move to the opposite directions, but consequently, the average effect could be that the density of states in the low binding energy region increases, although total energy is saved by the overall band movements.
Fig.~\ref{VB}(c) shows the comparison of corresponding EDC's in normal and AFM states.  One could see the features near $-0.6$~eV shift to lower binding energies, same as that in Fig. ~\ref{TD}(g). The features at high binding energies are smeared out, and one cannot resolve the band shift. Presumably, they would more towards the higher binding energy direction to save energy as shown in the calculations.

\section{Discussion and conclusion}

Previous ARPES results of  iron pnictides \cite{Yang,YiBaCo,HeNaFeAs} and FeSe thin films  on SrTiO$_3$ substrate \cite{Tan} show that electronic structure reconstruction rather than Fermi surface nesting drives the SDW transition. Theoretical studies on iron-based superconductors and their parent compounds suggest that the Hund's rule coupling is a key factor for the correlations and local moments \cite{Hund1, Hund2, Hund3}. The recent transport measurement on detwined FeTe suggests that Hund's rule coupling dominates magnetism in FeTe compound, which causes strong reconstruction and nematicity of the electronic structure \cite{JiangFeTe}. For FeTe, the system is largely characterized by a polaronic electronic structure \cite{zhangFeTe,liuFeTe}, and the bicollinear antiferromagnetic  transition there corresponds to large energy ($\sim$0.6~eV) and momentum (over the entire BZ)  scale spectral weight transfer.
For TaFe$_{1.23}$Te$_3$,  its large local moment of 2~$\mu_B$/Fe  is comparable to that of FeTe \cite{DaiFeTe}. The difference is that  TaFe$_{1.23}$Te$_3$ possesses 2D Fermi surface and significant in-plane dispersions.
Therefore, the key electronic character here is the coexisting of itinerant and localized $3d$ states, and
Hund's rule coupling  between them would introduce the double-exchange ferromagnetism.
This actually gives a naturally interpretation of the magnetic order in TaFe$_{1.23}$Te$_3$, as
the in-plane magnetic order  is determined by the competition between double exchange ferromagnetism and superexchange antiferromagnetism \cite{Hund2}. The chain-like structure of TaFe$_{1.23}$Te$_3$ significantly blocks the inter-chain superexchange between the  in-plane localized spins as they are too far away, while the inter-chain hoppings of the itinerant electrons can still be mediated by the intermediate Ta (nonmagnetic $5d^1$) structure.
 Therefore, the in-plane    FM coupling wins over the AFM one, resulting in the in-plane FM order
 shown in Fig.~1(b). On the other hand, the intra-ladder Fe-Fe or Fe-Te-Fe structure and the distance between planes [Fig.~1(a)] are similar to that of FeTe, and the direct FM exchange within a ladder and the inter-plane AFM superexchange  may work similarly as in FeTe.  Therefore, these conspire  the so called A-type AFM order in TaFe$_{1.23}$Te$_3$,  as in the language used for manganites.


To summarize, we have studied the electronic structure of TaFe$_{1.23}$Te$_3$ and its reconstruction across the AFM transition  with ARPES. The quasi-two-dimensional Fermi surface and band structure were observed, which qualitatively agree with the band calculations, and  indicate  sizable inter-ladder interactions within the plane. More importantly, our results suggest that TaFe$_{1.23}$Te$_3$ is the second kind of novel quantum materials beyond the parent compounds of iron-based superconductors, whose AFM transition directly correlates with the electronic structure reconstruction at high binding energies, instead of the traditional Fermi surface nesting. The commonalities among TaFe$_{1.23}$Te$_3$, FeTe, and other parent compounds of iron-based superconductors suggest that they are just the different manifestations out of the competition among the same set of physical ingredients, such as the Hund's rule coupling and antiferromagnetic superexchange interactions between localized spins \cite{Hund2}.  Particularly,
this spin ladder system would  provide a simpler  testing ground   to study the essence of magnetism in  iron pnictides and chalcogenides, which may even enable an analytical solution of the related theoretical model.

\section{Acknowledgement}

We are grateful to  Prof. Jiangping Hu and Prof. Hua Wu for helpful discussions. This work is supported in part by the National Science Foundation of China, and National Basic Research Program of China (973 Program) under the grant Nos. 2012CB921400, 2011CB921802 and 2011CBA00112.

\end{document}